\documentclass[12pt,reqno,a4paper,twoside]{article}
\usepackage{amsmath,amsthm,amstext,amscd,amssymb,euscript}
\usepackage{epsf}
\renewcommand{\phi}{\varphi}

\renewcommand{\subset}{\subseteq}
\renewcommand{\emptyset}{\varnothing}

\def\limn{\lim_{n\to\infty}}
\def\disagree{{\nleftrightarrow}}

\renewcommand{\Pr}{\mathbb P}

\def\1{ {\mathit{1} \!\!\>\!\! I} }

\makeatletter
\@addtoreset{equation}{section}
\makeatother

\newtheorem{ittheorem}[]{Theorem}
\newtheorem{itlemma}{Lemma}
\newtheorem{itproposition}{Proposition}
\newtheorem{itdefinition}{Definition}
\newtheorem{itremark}{Remark}

\newenvironment{theorem}{
\begin{ittheorem}}{\end{ittheorem}}

\newenvironment{lemma}{
\begin{itlemma}}{\end{itlemma}}

\newenvironment{proposition}{
\begin{itproposition}}{\end{itproposition}}

\newenvironment{definition}{
\begin{itdefinition}}{\end{itdefinition}}

\newenvironment{remark}{
\begin{itremark}}{\end{itremark}}

\newcommand{\beq}{\begin{eqnarray}}
\newcommand{\eeq}{\end{eqnarray}}

\newcommand{\be}{\begin{equation}}
\newcommand{\ee}{\end{equation}}

\newcommand{\bl}{\begin{lemma}}
\newcommand{\el}{\end{lemma}}

\newcommand{\br}{\begin{remark}}
\newcommand{\er}{\end{remark}}

\newcommand{\bt}{\begin{theorem}}
\newcommand{\et}{\end{theorem}}

\newcommand{\bd}{\begin{definition}}
\newcommand{\ed}{\end{definition}}

\newcommand{\bp}{\begin{proposition}}
\newcommand{\ep}{\end{proposition}}

\newcommand{\bc}{\begin{corollary}}
\newcommand{\ec}{\end{corollary}}

\newcommand{\bpr}{\begin{proof}}
\newcommand{\epr}{\end{proof}}

\newcommand{\bi}{\begin{itemize}}
\newcommand{\ei}{\end{itemize}}

\newcommand{\ben}{\begin{enumerate}}
\newcommand{\een}{\end{enumerate}}

\newcommand{\Z}{\mathbb Z}
\newcommand{\R}{\mathbb R}
\newcommand{\N}{\mathbb N}

\newcommand{\Q}{\mathbb Q}
\newcommand{\E}{\mathbb E}
\newcommand{\T}{\mathbf T}
\newcommand{\M}{\mathcal M}
\newcommand{\ret}{\mathbf R}
\newcommand{\veee}{\mathbb V}
\newcommand{\veek}{\ensuremath{\mathcal{V}}}
\newcommand{\gee}{\ensuremath{\mathcal{G}}}
\newcommand{\pee}{\ensuremath{\mathbb{P}}}

\newcommand{\ce}{\ensuremath{\EuScript{C}}}

\newcommand{\weee}{\ensuremath{\mathbf W}}
\newcommand{\meee}{\ensuremath{\mathbf M}}
\newcommand{\fe}{\ensuremath{\mathcal{F}}}

\newcommand{\la}{\ensuremath{\Lambda}}
\newcommand{\si}{\ensuremath{\sigma}}
\newcommand{\epsi}{\ensuremath{\epsilon}}
\newcommand{\gap}{\ensuremath{\Delta}}
\newcommand{\QED}{\hspace*{\fill}$\Box$\medskip}
\newcommand{\prodpee}{\ensuremath{\mathbb{P}\otimes\mathbb{P}}}
\newcommand{\prodE}{\ensuremath{\mathbb{E}\times\mathbb{E}}}

\def\now{
\ifnum\time<60       
          12:\ifnum\time<10 0\fi\number\time am 
          \else
            \ifnum\time>719\chardef\a=`p\else\chardef\a=`a\fi 
          \hour=\time
          \minute=\time 
          \divide\hour by 60 
          \ifnum\hour>12\advance\hour by -12\advance\minute by-720 \fi
          \number\hour:%
          \multiply\hour by 60 
          \advance\minute by -\hour
          \ifnum\minute<10 0\fi\number\minute\a m\fi}           
\newcount\hour 
\newcount\minute
\numberwithin{equation}{section}         

\theoremstyle{remark}

\begin{document}

\title{{\bf Occurrence, repetition and matching
of\\ patterns in the low-temperature\\ Ising model}}

\author{
J.-R. Chazottes
\footnote{CPhT, CNRS-Ecole Polytechnique, 91128 Palaiseau Cedex,
  France, jeanrene@cpht.polytechnique.fr}\\
F.\ Redig
\footnote{Faculteit Wiskunde en Informatica, Technische Universiteit Eindhoven, Postbus 513,
5600 MB Eindhoven, The Netherlands}\\
}

\maketitle

\begin{abstract}
We continue our study of the exponential law for occurrences and
returns of patterns in the context of Gibbsian random fields.
For the low-temperature plus-phase of the Ising model, we
prove exponential laws with error bounds for occurrence,
return, waiting and matching times.
Moreover we obtain a Poisson law for the number of occurrences
of large cylindrical events and a Gumbel law for the maximal
overlap between two independent copies. As a by-product,
we derive precise fluctuation results for the logarithm
of waiting and return times.
The main technical tool we use, in order to control
mixing, is disagreement percolation.
\end{abstract}

\bigskip


\section{Introduction}

The study of occurrence and return times
for highly mixing random fields has been
initiated by Wyner, see \cite{wyner}. In the
context of stationary {\em processes}, there is a vast
literature on exponential laws with error bounds
for $\alpha,\phi,\psi$-mixing processes, 
see e.g. \cite{AG} for a recent overview.
In the last four years, very precise
results were obtained by Abadi \cite{miguel}.
The advantage of his approach is that it gives sharp
bounds on the error of the exponential approximation
and it holds for {\em all} cylindrical events. Moreover,
it can be generalized to a broad class of random fields,
see \cite{acrv} for the case of Gibbsian random fields
in the Dobrushin uniqueness regime (high temperature).

Low-temperature Gibbsian random fields do not share the mixing
property of the Dobrushin uniqueness regime, i.e.
they are not (non-uniformly) $\phi$-mixing. 
So far, no results on exponential laws have been proved in this context.
To study these questions for Gibbsian random fields
at low temperature, the Ising model is a natural candidate to begin with.
The typical picture of the low-temperature plus-phase of this model
is a sea of plus spins with exponentially damped islands of minus spins.
Therefore decay of correlations of local observables can be
estimated using the technique of disagreement percolation as
initiated in \cite{vdberg} and further exploited in \cite{berg}. 

In this paper we prove the exponential law with error
bounds for occurrences and returns of cylindrical
events for the low-temperature plus-phase of the
Ising model.
As an application we also obtain the exponential
law with error bounds for waiting and matching times.
These results can then be further exploited to obtain a Poisson law
for the number of occurrences of cylindrical events
(the Poisson law for the number of large contour has been obtained
in \cite{ferrari} in the limit of zero temperature).
We also derive a `Gumbel law' for the maximal overlap
(in the spirit of \cite{karlin}) between
two independent copies of the low-temperature Ising model.
Other applications are strong approximations and large deviation
estimates of the logarithm of waiting and return times.
Our results are based upon disagreement percolation estimates and
are not limited to the Ising model only. However in this paper we
restrict to this example for the sake of simplicity. 

The paper is organized as follows. In Section \ref{setup}
we introduce basic notations, define occurrence and return
times, and collect the mixing results at low temperature
based on disagreement percolation.
In Section \ref{results} we state our results.
Section \ref{proofs} is devoted to proofs.

\section{Notations, definitions}\label{setup}

\subsection{Configurations, Ising model}

We consider the low-temperature plus-phase
of the Ising model on $\Z^d$, $d\geq 2$. This is a probability measure $\pee^+_\beta$ on
lattice spin configurations
$\si\in\Omega = \{+,-\}^{\Z^d}$, defined as the weak limit as $V\uparrow\Z^d$
of the following finite volume measures:
\be\label{rat}
\pee^+_{V,\beta}(\si_V) = 
\exp\left(-\beta\sum_{<xy>\in V}
  \si_x\si_y-\beta\sum_{\begin{subarray}{c}<xy> : \\ x\in\partial V,\ y\notin V\end{subarray}}\si_x\right)
\Big/ Z_{V,\beta}^+
\ee
where $Z_{V,\beta}^+$ is the partition function.
In \eqref{rat} $<xy>$ denotes nearest neighbor bonds and $\partial V$ the inner boundary,
i.e. the set of those $x\in V$ having at least one neighbor $y\notin V$.
For the existence of the limit $V\uparrow \Z^d$ of $\pee_{V,\beta}^+$, see e.g. \cite{Geo}. 

For $\eta\in \Omega$ $, V\subset \Z^d$ we denote by
$\pee_{V,\beta}^\eta$ the corresponding finite volume
measure with boundary condition $\eta$:
\be
\nonumber
\pee^\eta_{V,\beta}(\si_V) = \exp\left(-\beta \sum_{<xy>\in V}
\si_x\si_y- \beta \sum_{\begin{subarray}{c}<xy> : \\ x\in\partial V,\ y\notin V\end{subarray}}
\si_x\eta_y\right)
\Big/ Z_{V,\beta}^{\eta}\,.
\ee
Later on, we shall omit the indices $\beta,+$ (in $\pee^+_\beta$)
referring to the inverse temperature and plus
boundary condition respectively.
We will choose $\beta >\beta_0 >\beta_c$, i.e., temperature below 
the transition point, such that a certain mixing condition,
defined in detail below, is satisfied.

Let $V_n\uparrow\Z_+^d$ be an increasing sequence of sets such that
\be
\nonumber
\lim_{n\to\infty}\frac{|\partial V_n|}{|V_n|}=0\, .
\ee
In view of a later application to large deviation estimates,
we need the following pressure function $q\mapsto P(q\beta)$, $q\in\R$:
\be\label{qpressure}
P(q\beta)=\lim_{n\to\infty} \frac1{|V_n|} \log 
\sum_{\si_{V_n}\in \{+,-\}^{V_n}} \exp \left(
-q\beta \sum_{<xy>\in V_n} \si_x \si_y
 \right) \,.
\ee
(See \cite{Geo} for the existence of $P(q\beta)$.)

\subsection{Patterns, occurrence, repetition and matching times}

A pattern supported on a set $V\subset \Z^d$
is a configuration $\si_V\in\{+,-\}^V$. Patterns
will be denoted by $A$. We will identify $A$
with its cylinder, i.e., with the set
$\{ \si\in\Omega: \si_V = A\}$, so that it
makes sense to write e.g. $\si\in A$.
For $x\in\Z^d$, $\theta_x$ denotes the shift
over $x$. For a pattern $A$ supported
on $V$, $\theta_x A$ denotes the pattern
supported on $V+x$ defined by $\theta_x A (y+x) =A(y)$, $y\in V$.
We observe that for any Gibbs measure, so in particular in our
context, we have the uniform estimate 
\beq\label{oursin}
\pee(\sigma_V = A)\leq e^{-\delta |V|}
\eeq
for some $\delta>0$ and all patterns $A$.

If $A$ is a pattern supported on $V$, and $W\subset\Z^d$
then we denote by $(A\prec W)$ the event that
there exists $x\in\Z^d$ such that $V+x\subset W$ and
such that $\si_{V+x}= \theta_x A$.
In words this means that the pattern $A$ appears in the
set $W$.

Let $\veee=(V_n)$ where $V_n\uparrow\Z_+^d, $ is such that
$\lim_{n\to\infty}\frac{|\partial V_n|}{|V_n|}=0$,
and $A_n$ a pattern supported on $V_n$. We define
\be
\nonumber
\T^\veee_{A_n} 
=\min\{ |V_k|: A_n\prec V_k\}\,. 
\ee
In words, this is volume of the first set $V_k$ in which
we can see the pattern $A_n$.

For $n\in\N$ let $\ce_n$ be $[0,n]^d \cap \Z^d$.
We denote for $x\in\Z^d$:
$C(x,n)= \ce_n +x$. 
For $x,y\in\Z^d$: $|x-y|= \max_{i=1}^d |x_i-y_i|$, and for subsets $A,B\subset\Z^d$:
$d(A,B)=\min_{x\in A, y\in B} |x-y|$.

For $\si\in\Omega$, $A$ a pattern supported on $V$, $W\supset V$,
we define
the number of occurrences of $A$ in $W$:
\be
\nonumber
N(A,W,\si)= \sum_{x\in W: V+x\subset W} I( \si_{V+x} = \theta_x A)\,.
\ee

For a sequence $V_n\uparrow\Z_+^d$, the return time is defined as follows:

\be
\nonumber
\ret_{\si_{V_n}} (\si ) = \min \{ |V_k|: N(\si_{V_n},V_k,\si)\geq 2\}\,.
\ee

Finally, for $\veee= V_n\uparrow \Z_+^d$, and $\si$, $\eta\in\Omega$, we
define the waiting time:

\be
\nonumber
\weee (V_n,\eta,\si) =\T^\veee_{\eta_{V_n}} (\si)\,.
\ee

We are interested in this quantity for
$\si$ distributed according to $\pee$ and $\eta$
distributed according to another ergodic
(sometimes Gibbsian) probability measure $\Q$ .

Finally, we consider `matching times', in view of studying
maximal overlap between two independent samples of $\pee$.
For $\si$, $\eta\in \Omega$, 
$$
\meee(V_n,\si,\eta)= \min\{ |V_k|: \exists x:\ V_n +x \subset V_k,\ \si_{V_n +x}=\eta_{V_n +x}\}\,.
$$
In words, this is the minimal volume of a set of type $V_k$ such that
inside $V_k$, $\si$ and $\eta$ match on a set of the form $V_n +x$.

In the sequel we will omit the reference to the sequence
$V_n$, in order not to overburden notation. In fact, proofs
will be done for $V_n=\ce_n= [0,n]^d\cap\Z^d$.
The generalization to $\veee$ is obvious provided that
the following two (sufficient) conditions are fulfilled:
\ben
\item $\lim_{n\to\infty}\frac{|\partial V_n|}{|V_n|}=0$;
\item There exists $c>0$ such that, for all $x$ with $|x|\geq 1$, $|(V_n+x) \Delta V_n|\geq c n$\,.
\een

\subsection{Mixing at low temperatures}

In \cite{acrv} we derived exponential laws
for hitting and return times under a mixing condition of
the type

\be\label{kwak}
\sup_{\si,\eta,\xi}|\pee_{V}^\eta (\si_W ) -\pee_V^\xi (\si_W)|
\leq |W| \exp (-c d(V^c,W))
\ee
usually called `non-uniform exponential $\phi$-mixing'.
This condition is of course not satisfied at low
temperatures since boundary conditions
continue to have influence. Take e.g. $W=\{ 0 \}$, $\eta \equiv +$,
$\xi\equiv -$, then for $\beta >\beta_c$ :
\be
\nonumber
\lim_{V\uparrow\Z^d} \pee_{V}^\eta (\si_0 =+) -\pee_V^\xi (\si_0=+)
= m^+_\beta >0
\ee
where $0<m^+_\beta = \int \si_0 d\pee (\si)$ is the magnetization. This
clearly contradicts (\ref{kwak}).
However, for local functions $f,g$ we do have an estimate
like
\be
\nonumber
\Big|\int f\ \theta_x g\ d\pee - \int f d\pee \int gd\pee\Big|
\leq C(f,g)\ e^{-c(\beta) |x|}\,.
\ee
The intuition here is that there can only be correlation
between two functions if the clusters containing
their dependence sets are finite (i.e. not contained in the sea of pluses) and intersect.
Since finite clusters are exponentially small (in diameter), we have exponential decay of correlations of
local functions.

This idea is formalized in the context of `disagreement percolation'.
To introduce this concept, we define a path $\gamma =\{ x_1,\ldots , x_n \}$, i.e.
a subset of $\Z^d$ such that $x_i$ and $x_{i-1}$ are neighbors for all $i=1,\ldots , n$.

More formally, for $W\subset V$ and $\eta$ and $\xi\in\Omega$, we have the following inequality:
\be\label{disa}
|\pee_V^\eta (\si_W) -\pee_V^\xi (\si_W)|
\leq 
|\partial W|\ \pee_V^\eta\otimes\pee_V^\xi (W\disagree \partial V)\,.
\ee
Here $(W\disagree \partial V)$ denotes the event of
those couples $(\si_1,\si_2)\in\Omega_V\times\Omega_V$
where there is `a path of disagreement'
$\gamma$ leading from $W$ to the boundary of $V$
such that $\si_1(x)\not=\si_2 (x)$ for all $x\in\gamma$.
Of course whether the probability of this
event under the measure $\pee_V^\eta\otimes\pee_V^\xi $
will be small depends on the distance
between $V$ and $W$ and on the chosen boundary
conditions $\eta,\xi$.
The estimate (\ref{disa}) as well as
the ideas of disagreement percolation can be found in
\cite{berg},\cite{ghm}.

On the top of inequality \eqref{disa}
we have the following estimate of \cite{BS}, see \cite{ghm}:
\be\label{burs}
\pee\otimes\pee (\partial W\disagree \partial V)
\leq e^{-c(\beta) d(W,\partial V)}
\ee
as soon as $\beta>\beta_0 (>\beta_c)$, and where $c(\beta)\to\infty$ as $\beta\to\infty$.

In the rest of the paper we always work with $\beta >\beta_0$,
so that we can apply (\ref{disa}), (\ref{burs}). We emphasize
that the next results are in fact valid not only for the Ising
model at low temperature but also for any Markovian random field
for which the above disagreement percolation estimates hold.

\section{Results}\label{results}

\subsection{Exponential laws}

\bt[Occurrence times]\label{main}
There exist $0<\Lambda_1\leq \Lambda_2 <\infty$, $c,c' >0$,
$0<\kappa<c$, such that for all patterns $A=A_n$ supported on $\ce_n$,
there exists $\lambda_A\in [\Lambda_1,\Lambda_2]$ 
such that for all $n$ and all $t < e^{\kappa n^d}$:
\be\label{mainequation}
\Big|\pee \left(\T_A \geq \frac{t}{\lambda_A \pee (A)}\right) -e^{-t}\Big|
\leq e^{-ct}e^{-c' n^d}.
\ee
\et

For return times we have
to restrict to "good patterns", i.e., patterns which
are not `badly self-repeating' in the following sense:
\bd\label{rouletterusse}
A pattern $A_n$ is called good
if for any $x$ with $|x|<n/2$, for the
cylinders we have
$A_n\cap\theta_x A_n=\emptyset$.
\ed
Good patterns have a return time at least
$(n/2+1)^d$ and as we will see later that this property guarantees
that the return time is actually of the
order $e^{cn^d}$.

The following lemma is proved in \cite{acrv} for
general Gibbsian random fields.

\bl\label{goodpatterns}
Let $\gee_n$ be the set of all good patterns.
There exists $c >0$ such that 
\be
\nonumber
\pee(\gee_n) \geq 1-e^{-c n^d}\,.
\ee
\el

We denote by $\pee (\cdot|A)$ the measure $\pee$ conditioned
on the event $A\prec \ce_n$.

\bt[Repetition time]\label{main2}
There exist $0<\Lambda_1\leq \Lambda_2 <\infty$, $c,c' >0$,
$0<\kappa<c$, such that for all {\em good} patterns $A=A_n$ supported
on $\ce_n$, there exists $\lambda_A\in [\Lambda_1,\Lambda_2]$
such that for all $n$ and all $t < e^{\kappa n^d}$:
\be\label{mainequation2}
\Big|\pee \left(\ret_A \geq \frac{t}{\lambda_A \pee (A)} \ \Big | A\right) -e^{-t}\Big|
\leq e^{-ct}e^{-c' n^d}.
\ee
\et

We have the following analogue of Theorem \ref{main} for matching
times.

\bt[Matching time]\label{match}
There exist $0<\Lambda_1\leq \Lambda_2 <\infty$, $c,c' >0$,
$0<\kappa<c$, such that for all patterns $A=A_n$ supported on $\ce_n$,
there exists $\lambda_A\in [\Lambda_1,\Lambda_2]$ 
such that for all $n$ and all $t < e^{\kappa n^d}$:
\be\label{matchequation}
\Big|\pee\!\otimes\!\pee \left((\si,\eta):\meee_n (\si,\eta)\geq
\frac{t}{\lambda_n \pee\!\otimes\!\pee(\si_{\ce_n}=\eta_{\ce_n})}\right) -e^{-t}\Big|
\leq e^{-ct}e^{-c' n^d}.
\ee
\et

\subsection{Poisson law}

Let $A=A_n$ be any pattern supported on $\ce_n$. For $t>0$,
let $C(t/\pee (A))$ be the maximal cube of the form $C_k= [0,k]^d\cap\Z^d$ such that
$|C_k|\leq t/\pee (A)$. Observe that 
\[
\frac{|C(t/\pee (A))|}{t/\pee (A)}\to 1
\]
as $n\to \infty$.
Define
\be\label{croquemonsieur}
N^{n}_t (\si)= N( A_n,C(t/\pee (A)),\si)\,.
\ee
Then we have

\bt\label{Poissonlaw}
If $\si$ is distributed according to $\pee$, and $A_n$ is a sequence
of good patterns, then
the processes $\{ N^{n}_t/\lambda_{A_n}: t\geq 0\}$ converge to a mean
one Poisson process $\{ N_t :t\geq 0\}$ weakly on path space,
where $\lambda_{A_n}$ is the parameter of Theorem \ref{main}.
\et

\subsection{Gumbel law}

To formulate the Gumbel law for certain extremes, we need
simply connected subsets $G_n$, $n\geq 1$, such that $|G_{n}|=n$
and $G_{n^d}=\ce_n$.
For instance, for $d=2$, $G_1=\{(0,0)\}$, $G_2=\{(0,0),(1,0)\}$,
$G_3=\{(0,0),(1,0),(1,1)\}$, $G_4=\{(0,0),(1,0),(1,1), (0,1)\}$, etc.

For $\eta\in\Omega$, define
\be\label{Gumbellaw}
\M_n(\eta ,\si) =\max \{ |G_k| : \exists x\in G_n \ \mbox{with}\ G_{k}+x\subset G_n
\ \mbox{and}\ \eta_{G_k+x}=\si_{G_k+x}\}
\ee
In words this is the volume of the maximal subset of the type $G_k$ on which
$\eta$ and $\si$ agree. We have the following

\bt\label{Gumbelthm}
For any $\eta\in\Omega$,
there exists a sequence $u_n\uparrow \infty$,
and constants $\lambda,\lambda',\nu,\nu'\in (0,\infty)$
such that for all $x\in\Z$
\beq\label{gumbel}
& & \min\{e^{-\lambda'e^{-\nu' x}}, e^{-\lambda e^{-\nu x}}\}
\leq \liminf_{n\to\infty}
\prodpee \left((\eta,\si):\M_n(\eta,\si) \leq u_n + x\right) 
\leq 
\nonumber\\
& & \limsup_{n\to\infty}
\prodpee \left((\eta,\si):M_n(\eta,\si)\leq u_n + x\right)
\leq  
\max\{e^{-\lambda'e^{-\nu' x}}, e^{-\lambda e^{-\nu x}}\}\,.
\eeq
\et

The fact that in the Gumbel law we only have a lower and an
upper bound is due to the discreteness of the $\meee_n(\si,\eta)$.
This situation can be compared to the study of the maximum of
independent geometrically distributed random variables, see for instance \cite{galambos}.

\br
Notice that in Theorem \ref{Gumbelthm} we study the
maximal matching between two configurations on
a specific sequence of supporting sets $G_n$. Since in
the low-temperature plus-phase we have percolation
of pluses, the same theorem would of course
not hold for the cardinality of the maximal
connected subset of $\ce_n$ on which $\eta$ and $\si$ agree
because
the latter subset occupies a fraction of the volume
of $\ce_n$.
\er

\subsection{Fluctuations of waiting, return and matching times}

We denote by $s(\Pr)$ the entropy of $\Pr$ defined by
$$
s(\Pr)=\lim_{n\to\infty}-\frac1{n^d}\sum_{A_n\in \{+,-\}^{\ce_n}}\Pr(A_n)\log\Pr(A_n)\, .
$$

The next result (proved in Subsection
\ref{proof-corollary-OW}) shows how the repetition of
typical patterns allows to compute the entropy from a single `typical'
configuration.

\bt\label{corollary-OW}
There exists $\epsilon_0 >0$ such that
for all $\epsilon>\epsilon_0$
\be\label{strong-approximation}
-\epsilon\log n \leq\log\left[\ret_{\si_{\ce_n}}(\sigma)\
\Pr(\sigma_{\ce_n})\right]\leq \log\log n^{\epsilon}
\quad\textup{eventually}\; \Pr\!-\!\textup{almost surely}.
\ee
In particular,
\be\label{NUEMOW}
\lim_{n\to\infty}\frac{1}{n^d}\log\ret_{\si_{\ce_n}}(\sigma)=s(\Pr)\quad\Pr-\textup{almost
surely}\,.
\ee
\et

Note that \eqref{NUEMOW} is a particular case of the result by Ornstein
and Weiss in \cite{OW} 
where $\Pr$ is only assumed to be ergodic. Under our assumptions, we get
the more precise
result \eqref{strong-approximation}.

\br
It follows immediately from \eqref{strong-approximation} that the sequence
\textup{(}$\log\ret_{\si_{\ce_n}}(\si)/n^d$\textup{)} satisfies the central limit theorem if and
only if \textup{(}$-\log\pee(\si_{\ce_n})/n^d$\textup{)} does. However, in the low-temperature
regime, we are not able to prove the central limit theorem for 
\textup{(}$-\log\pee(\si_{\ce_n})/n^d$\textup{)}.
\er

Suppose that $\eta$ is a configuration randomly chosen
according to an ergodic random field $\mathbb{Q}$ and, independently, $\sigma$ is
randomly chosen according to $\Pr$.
We denote by
$s(\mathbb{Q}|\Pr)$ the relative entropy density
of ${\mathbb Q}$ with respect to $\Pr$, where
$$
s(\mathbb{Q}|\Pr)=\lim_{n\to\infty}\frac1{n^d}
\sum_{A_n\in\{+,-\}^{\ce_n}}\mathbb{Q}(A_n)\log\frac{\mathbb{Q}(A_n)}{\Pr(A_n)}\,\cdot
$$

We have the following result (proved in Subsection \ref{wtproof}):

\bt\label{waiting-time}
Assume that $\mathbb{Q}$ is an ergodic random field. Then 
there exists $\epsilon_0 >0$ such that
for all $\epsilon>\epsilon_0$
\be\label{dracula}
-\epsilon\log n \leq\log\left(\weee(\ce_n,\eta,\sigma))\ 
\Pr(\eta_{\ce_n})\right)\leq \log\log n^\epsilon
\ee
for $\mathbb{Q}\otimes\Pr$-eventually almost every $(\eta,\sigma)$.
In particular
\be\label{ASW}
\lim_{n\to\infty}\frac{1}{n^d}\log
\weee(\ce_n,\eta,\sigma)=s(\mathbb{Q})+s(\mathbb{Q}|\Pr)
\quad\mathbb{Q}\otimes\Pr-\textup{a.s}\, .
\ee
\et

\br
If in \eqref{ASW} we choose $\mathbb{Q}=\pee^-$, the low-temperature minus-phase, we conclude that the time
to observe a pattern typical for the minus phase in the plus phase, is 
equal to the time to observe a pattern typical for the plus phase, at the logarithmic
scale.
\er

The next theorem is proved in Subsection \ref{LD-proof}.
\bt\label{LD-waiting}
For all $q\in\R$ the limit
\be
\nonumber
\mathcal{W} (q)= \lim_{n\to\infty} \frac{1}{n^d} \log \int \weee(\ce_n,\eta,\sigma)^{q}\
d\Pr\!\otimes\!\Pr (\eta,\si)
\ee
exists and equals
\be\label{w-cases}
\mathcal{W}(q)= \begin{cases} 
P((1-q)\beta) + (q-1)P(\beta) & \mbox{ for }q\ge -1\\
P(2\beta) - 2P(\beta) &\mbox{ for }q<-1
\end{cases}
\ee
where $P$ is the pressure defined in \eqref{qpressure}.
\et

From this result, it follows that the sequence
($\frac{1}{n^d}\log\weee(\ce_n,\eta,\sigma)$)
satisfies a generalized large deviation principle in the sense 
of Theorem 4.5.20 in \cite{DZ}. The differentiability of $q\mapsto P(q\beta)$
would imply a full large deviation principle. 

\br
A more general version of Theorem \ref{LD-waiting} can be easily
derived: The measure $\Pr \otimes \Pr$ can be replaced
by the measure ${\mathbb Q}\otimes \Pr$ where ${\mathbb Q}$ is any Gibbsian
random field (without any mixing assumption). Of course formula (\ref{w-cases}) has to be
properly modified (see \cite{acrv}).
\er

For the matching times, we have the following analogue
of Theorem \ref{waiting-time} (see Subsection \ref{pape}):

\bt\label{matching-time}
There exists $\epsilon_0 >0$ such that
for all $\epsilon>\epsilon_0$
\be\label{vampire}
-\epsilon\log n \leq\log\left(\meee(\ce_n,\eta,\sigma)\ 
\pee\!\otimes\!\pee(\si_{\ce_n}=\eta_{\ce_n})\right)\leq \log\log n^\epsilon
\ee
for $\pee\!\otimes\!\pee$-eventually almost every $(\eta,\sigma)$.
In particular
\be\label{ASM}
\lim_{n\to\infty}\frac{1}{n^d}\log
\meee(\ce_n,\eta,\sigma)= \mathcal{W}(-1)
\quad\pee\!\otimes\!\pee-\textup{a.s}\, .
\ee
\et

\section{Proofs}\label{proofs}

From now on, we write $A$ for $A_n$ to alleviate notations.
(Therefore $A$ is understood to be a pattern supported on $\ce_n$.)

\subsection{Positivity of the parameter}

The following lemma is the analogue of Lemma 4.3 in \cite{acrv}. 

\bl[The parameter]\label{pospar}
There exist strictly positive constants $\Lambda_1,
\Lambda_2$ such that
for any integer  $t$  with $t\Pr(A)\le 1/2$, one has 
$$
\Lambda_1\le \lambda_{A,t} := -\frac { \log\Pr( \T_A>t)}{t \Pr(A)}\le \Lambda_2\,.
$$
\el

\bpr
We proceed by estimating the second moment of the random variable
$N( A, \ce_k, \si)$, where $k=\lfloor t^{1/d}\rfloor$. We have
\be
\nonumber
\E (N( A, \ce_k, \si))^2 = \sum_{x,y: x+\ce_n\subset \ce_k, y+\ce_n\subset \ce_k} \pee( \theta_x A\cap\theta_y A).
\ee
We split the
sum in three parts:
$I_1 =\sum_{x=y}$, $I_2= \sum_{x\not= y, |x-y|\leq\Delta}$, $I_3=
\sum_{x\not= y, |x-y| >\Delta}$, where $\Delta>0$ will be specified
later on.

We now estimate $I_1$, $I_2$ and $I_3$.
The quantities $I_1$ and $I_2$ are estimated as in \cite{acrv}.
For $I_1$ we have:
\[
I_1= (k+1)^d \pee (A).
\]
For $I_2$, using the Gibbs property \eqref{oursin} and $d\geq 2$:
\[
I_2\leq (k+1)^d\Delta^d e^{-\delta n} \pee (A).
\]
Only the third term involves the disagreement percolation estimate. 
\beq
\nonumber
&&I_3 - (k+1)^{2d}\pee (A)^2 \nonumber\\  &\leq& 
\nonumber
\sum_{x\not= y, |x-y|> \Delta} \pee (A)\ |\pee( \si_{C(x,n)} =A|\si_{C(y,n)} =A) - \pee (A)|.
\eeq
Denote by $C'_{x,\Delta,n}$ the set of those sites
which are at least at lattice distance $\Delta +1$ away
from $C(x,n)$, and $C^\Delta(x,n)$ the complement of
that set. Then we have for $|x-y|>\Delta$:
$$
|\pee( \si_{C(x,n)} =\theta_x A|\si_{C(y,n)} =\theta_y A) - \pee (A)|=
$$
$$
\left|\int\int
\left(\pee (\si_{C(x,n)}=\theta_x A|\eta_{C'_{x,\Delta,n}})
-
\pee (\si_{C(x,n)}=\theta_x A|\xi_{C'_{x,\Delta,n}})\right)
d\pee (\eta|\si_{C(y,n)}=\theta_y A)d\pee (\xi)\right|
$$
$$
\leq 
\int \int
\pee_{C^\Delta(x,n)}^{\eta}\otimes\pee_{C^\Delta(x,n)}^\xi
\left( C(x,n)\disagree \partial C^\Delta(x,n)\right)
d\pee (\eta|\si_{C(y,n)} = \theta_x A)d\pee (\xi )
$$
$$
\leq 
\frac{1}{\pee (A)}\ \pee\otimes\pee
\left( C(x,n)\disagree \partial C^\Delta(x,n)\right)
\leq 
\frac{1}{\pee (A)}\ |\partial C(x,n)|\ e^{-d(C(x,n),\partial C^\Delta(x,n))}
$$
$$
\leq 
e^{-c n^{d+1} + c' n^d}\leq e^{-\tilde{c} n^{d+1}}
$$
where in the last step we made the choice $\Delta= \Delta_n = n^{d+1}$.
Using the second moment estimate (Lemma 4.2 in \cite{acrv}) and proceeding as in the proof of
Lemma 4.3 in \cite{acrv}, we obtain the inequality
\beq
\frac{\pee (\T_A\leq t)}{t\pee (A)}
&\geq&
\frac{1}{1+ e^{-\delta n}\Delta^d + t\pee (A) + e^{-c n^{d+1}} t/\pee (A)}
\nonumber\\
&\geq &
\frac{1}{1+ C_1 + 1/2 + C_2}
\nonumber
\eeq
where
\[
C_1=\sup_n n^{d(d+1)}\ e^{-\delta n} <\infty,\quad
C_2 = \sup_{A}\sup_{t\leq 1/(2\pee (A))} e^{-c n^{d+1}} t/\pee (A) <\infty\,.
\]
The upper bound is derived as in the high temperature case, see \cite{acrv}.
\epr

\subsection{Iteration lemma and proof of Theorem \ref{main}}\label{marx}

This is the analogue of Lemma 4.4 in \cite{acrv}. 

We consider $k$ mutually disjoint cubes $C_i$ such that
$|C_i|= f_A= (\lfloor \pee (A)^{-\theta/d} \rfloor +1)^d$, where
$0<\theta<1$ is fixed. The essential point is to make precise
the approximation
of $\pee (A\nprec \cup_{i=1}^k C_i)$ by $\pee (A\nprec C_1)^k$. 

For a cube $C_i$ we denote by $C_i^{\Delta'}\subset C_i$ the largest
cube inside $C_i$ with the same midpoint as $C_i$
and such that the boundary $\partial C_i$
is at least at lattice distance $\Delta'$ away from $C_i^{\Delta'}$,
where $\Delta'=\Delta'(n,t)> n^{d+1}$ will be fixed later.
We have
$$
\pee \left(A\nprec \cup_{i=1}^k C_i\right)
=
$$
$$
\pee (A\nprec C_1|A\nprec C_2\cap A\nprec C_3\cap\cdots \cap A\nprec C_k)
\pee(A\nprec C_2\cap A\nprec C_3\cap\cdots \cap A\nprec C_k)=
$$
$$
\left(\pee (A\nprec C^{\Delta'}_1|A\nprec C_2\cap A\nprec C_3\cap\cdots\cap
  A\nprec C_k)+
\epsilon_1\right)\pee(A\nprec C_2\cap A\nprec C_3\cap\cdots \cap A\nprec C_k)=
$$
$$
\left( \pee (A\nprec C_1^{\Delta'}) +\epsilon_1 +\epsilon_2\right)
\pee(A\nprec C_2\cap A\nprec C_3\cap\cdots \cap A\nprec C_k)=
$$
$$
\left( \pee (A\nprec C_1) +\epsi_1 +\epsi_2 +\epsi_3\right)
\pee(A\nprec C_2\cap A\nprec C_3\cap\cdots\cap A\nprec C_k)\,.
$$
We now start to estimate the errors $\epsi_i$.
For the first one:
\beq
|\epsi_1|
&\leq &
\pee (A\nprec C_1^{\Delta'} \cap A\prec C_1|A\nprec C_2\cap A\nprec
C_3\cap\cdots \cap A\nprec C_k)
\nonumber\\
&\leq &
\Delta' f_A^{(d-1)/d} \pee (A)\ e^{c n^{d-1}}\,.
\nonumber
\eeq
In the last step, the factor $e^{c n^{d-1}}$ arises
by removing the conditioning and using the following general property
of Gibbs measures:
\be
\nonumber
\sup_{\eta,\xi}\frac{ \pee (\si_{\ce_n} = A|\eta_{\ce_n^c})}
 {\pee (\si_{\ce_n} = A|\xi_{\ce_n^c})}
\leq e^{c n^{d-1}}\,.
\ee
For $\epsi_2$ we use the disagreement percolation estimate,
as in the proof of Lemma \ref{pospar}:
\be
\nonumber
|\epsi_2|
\leq 
\frac{ \pee\otimes\pee (C_1^{\Delta'} \disagree \partial C_1)}
{\pee (A\nprec C_2\cap A\nprec C_3\cap \cdots \cap A\nprec C_k)}
\leq e^{-c_1\Delta'} e^{c_2 n^d} \leq e^{-c n^{d+1}}
\nonumber
\ee
where $c_1,c_2,c>0$. Finally, proceeding as in the estimation of
$\epsi_1$, we get 
\be
\nonumber
\epsi_3 \leq \Delta' f_A^{(d-1)/d} \pee (A)
\ee
where now the boundary factor $e^{cn^{d-1}}$ is absent
since we do not have a conditioned measure.
Let
\be
\nonumber
\alpha_{k-p}= \pee (A\prec \cup_{i=p+1}^k C_i).
\ee
We obtain the recursion inequality:
\be
\nonumber
\alpha_k \leq (\alpha_1+\epsi_1+\epsi_3)\alpha_{k-1} +\epsi
\ee
where $\epsi\leq e^{-c n^{d+1}}$.
Following the lines of the proof of Lemma 4.4 in \cite[formula (38)]{acrv}
this gives
$$
\alpha_k- \alpha_1^k \leq
$$
$$
k\Bigl(2\Delta' f_A^{(d-1)/d} \pee(A) e^{cn^{d-1}}
\Bigr)
\Bigl(
\pee (A\nprec C_1) + 2\Delta' f_A^{(d-1)/d} \pee(A) e^{cn^{d-1}}
\Bigr)^{k-1}
+ k \epsi =:
\textup{I}+\textup{II}\,.
$$
Now, fix $f_A= \pee (A)^{-\theta}$, $\Delta'= t n^{d+1}$ and $k= \lfloor\frac{t}{\pee (A) f_A}\rfloor$.
Then we have
\[
\textup{I}\leq  t e^{-c n^d}\,.
\]
and
\[
\textup{II} \leq t e^{-ct n^{d+1}}\, .
\]
Therefore, as long as $t< e^{\kappa n^d}$ with $\kappa < c$, we have
\be
\nonumber
\alpha_k -\alpha_1^k \leq e^{-c' n^d}e^{-ct}\,.
\ee
The lower bound
\be
\nonumber
\alpha_k -\alpha_1^k \geq e^{-c' n^d}e^{-ct}
\ee
is obtained analogously.
At this stage, one can repeat the proof of \cite{acrv} to obtain 
\eqref{mainequation} in Theorem \ref{main}.
\QED

\subsection{Return time}

Let $C=\ce_{f_A^{1/d}}$ where $f_A=(\lfloor \pee
(A)^{-\theta/d} \rfloor +1)^d$. (Notice that $\ce_n\subset C$ as long as $n$ is large enough.)
For a pattern $A=A_n$ and a configuration $\si\in\Omega$
such that $\si_{\ce_n}= A$ we write $A\prec^* C$ for the event that $A$ appears at least twice
$C$ and $A\nprec^* C$ is the event
that $A$ occurs in $C$ only on $\ce_n$, i.e.,
the number of occurrences is equal to one.

In order to repeat the iteration lemma for pattern
repetitions, we first prove the following lemma.
\bl\label{goodlemma}
Let $A=A_n$ be a good pattern, then there exists $c >0$
such that for the cube $C=\ce_{f_A^{1/d}}$ where $f_A=(\lfloor \pee (A)^{-\theta/d} \rfloor +1)^d$,
we have
\be
\nonumber
\left|\pee( A\nprec^* C | A)-\pee(A\nprec C)\right|\leq e^{-c n^d}\,.
\ee
\el
\bpr
Since $A$ is good, $A$ does not appear in any
cube $\theta_x \ce_n$ for $|x| <n/2$. 
We will introduce a gap $\Delta$ with a $n$-dependence
to be chosen later on.
Denote
by $\ce_n^{\Delta}$ the minimal cube containing $\ce_n$ such
that its boundary is at distance at least $\Delta$
from $\ce_n$. We have
\beq
\nonumber
|\pee(A\nprec^* C|A)-
\pee(A\nprec^* C\setminus \ce^{\Delta}_n|A)|
&\leq &
\pee(A\prec \ce^{\Delta+n+1}_n\setminus \ce_{n/2}|A)
\nonumber\\
&\leq &
(\Delta+n+1)^d e^{-cn^d}\,.
\nonumber
\eeq
To get the last inequality, remark that
\be\label{kots}
\pee(A\prec \ce^{\Delta+n+1}_n\setminus \ce_{n/2}|A)
\leq 
|\ce^{\Delta+n+1}_n\setminus \ce_{n/2}|\ \sup_{V:|V|>(n/2)^d}\sup_{B
  \in \Omega_V}\sup_{\eta\in\Omega}\pee (B|\eta_{V^c}) 
\ee
since $|\theta_x \ce_n\setminus \ce_n|> (n/2)^d$ for $|x|\geq n/2$.
The rhs of (\ref{kots}) is bounded by $e^{-c n^d}$ by the
Gibbs property \eqref{oursin} and the fact that a conditioning can at
most cost a factor $e^{c n^{d-1}}$.
Now we can use the mixing property to obtain
\be
\nonumber
|\pee(A\nprec^* C\setminus \ce^{\Delta}_n|A)-
\pee(A\nprec C\setminus \ce^{\Delta}_n)|
\leq
e^{-c_1 \Delta} e^{c_2 n^d} f_A^{(d-1)/d}
\ee
and finally,
\be
\nonumber
|\pee (A\nprec C)-\pee(A\nprec C\setminus \ce^{\Delta}_n)|
\leq {\Delta} f_A^{(d-1)/d} \pee (A)
\ee
which yields the statement of the lemma by
choosing $f_A = (\lfloor \pee (A)^{-\theta/d}\rfloor +1)^d$ and $\Delta=n^{d+1}$.
\epr
We can now state the analogue of the iteration lemma
for pattern repetitions.
\bl
Let $A=A_n\in\gee_n$ be a good pattern. Let $C_i$, $i=1,\ldots,k$, be a collection
of disjoint cubes of volume $f_A$ such that $C_1=\ce_{f_A^{1/d}}$.
We have the following
estimate:
\beq
&&\pee(A\nprec^* \cup_{i=1}^k C_i|A) - \left[\pee (A\nprec C_1)\right]^k
\nonumber\\
&\leq &
k\Bigl(2\Delta f_A^{(d-1)/d} \pee(A) e^{cn^{d-1}}
\Bigr)
\Bigl(
\pee (A\nprec C_1) + 2\Delta f_A^{(d-1)/d} \pee(A) e^{cn^{d-1}}
\Bigr)^{k-1}
\nonumber\\
&+ &
k e^{-c\Delta} + e^{-cn^d} \pee (A\nprec C_1)^{k-1}\,.
\nonumber
\eeq
\el
\bpr
Start with the following identity:
\be\label{lu}
\pee(A\nprec^* \cup_{i=1}^k C_i|A)
=\frac{ \pee(A \cap A\nprec^* C_1\cap A\nprec C_2\cap \cdots \cap A\nprec C_k)}
{\pee (A)}\ \cdot
\ee
We can proceed now as in the proof of the iteration lemma
to approximate the rhs of (\ref{lu}) by
\be
\nonumber
\Pi_k= \frac{\pee (A\cap A\nprec^* C_1)}{\pee (A)}\
\pee (A\nprec C_2)\cdots \pee (A\nprec C_k)
\ee
at the cost of an error $\epsi$ which can be estimated by
\be
\nonumber
\epsi
\leq 
k\Bigl(2\Delta f_A^{(d-1)/d} \pee(A) e^{cn^{d-1}}
\Bigl(
\pee (A\nprec C_1) + 2\Delta f_A^{(d-1)/d} \pee(A)
e^{cn^{d-1}}\Bigr)^{k-1} + k e^{-c\Delta}\ .
\ee
Now, to replace $\Pi_k$ by $\pee (A\nprec C_1)^k$, use
Lemma \ref{goodlemma} to conclude
that this replacement induces an extra error which is at most
\be
e^{-cn^d} \pee (A\nprec C_1)^{k-1}\,.
\ee
The lemma is proved.
\epr

\subsection{Matching time}

In order to prove the exponential law (\ref{match}) for matching times, we first remark that
for cylinders $A_n$ defined on $\Omega\times\Omega=(\{+,-\}\times\{+,-\})^{\Z^d}$, we have the analogue of
Theorem \ref{main} under the measure $\pee\otimes\pee$ with the same proof. 
Indeed, a typical configuration drawn from $\pee\otimes\pee$ is a sea of
$(+,+)$ with exponentially damped islands of non $(+,+)$.
We now generalize the statement of Theorem \ref{main} to the $\fe_n$ measurable events
that we need (which are not cylindrical).

\bl
Suppose $E_n=\{ (\si,\eta):\si_x=\eta_x,\ \forall x\in \ce_n\}$.
Theorem \ref{main} holds with $A_n$ replaced by $E_n$ and $\pee$ replaced by $\prodpee$.
\el

\bpr
Clearly, the analogue of the iteration lemma does not pose any new problem.
The main point is to prove the non-triviality of the parameter, i.e.,
the analogue of Lemma \ref{pospar}. In order to obtain this, we have to estimate
the second moment of 
\[
N^k_{E_n} = \sum_{x:\ce_n+x\subset C_k} I(\theta_x E_n)
\]
under $\prodpee$.
As before we split
\be\label{wif}
\prodE (N^k_{E_n})^2 \leq I_1 +I_2 +I_3
\ee
where
$I_1= \sum_{x=y} \prodpee (E_n)\leq (k+1)^d \pee (E_n)$,
$I_2 =\sum_{x\not= y, |x-y|\leq \Delta} \prodpee (\theta_x E_n\cap \theta_y E_n)$
and
$I_3=\sum_{x\not= y, |x-y|> \Delta} \prodpee (\theta_x E_n\cap \theta_y E_n)$.
The only problematic term here is $I_2$. As in the proof for cylindrical events,
we will use the Gibbs property, and prove first the existence of $1>\delta>0$ such that
\be\label{waf}
\delta \leq \prodpee (\si_x=\eta_x| (\si,\eta)_{\Z^d\setminus\{x\}}) \leq 1-\delta\,.
\ee
We now further estimate
\beq
\nonumber
\prodpee (\si_x=\eta_x|
(\si,\eta)_{\Z^d\setminus\{x\}})= \sum_{\epsi= +,-} \pee(\si_x=\epsi|\si)\pee(\eta_x=\epsi|\eta)\\
\label{woef}
\leq \sup_{\si,\eta}\left[ \pee(+|\si)\pee(+|\eta) + (1-\pee(+|\si))(1-\pee(+|\eta))\right]\,.
\eeq
Since by the Gibbs property $0<\zeta<\pee(+|\eta)<1-\zeta < 1$, we can bound (\ref{woef}) by
\[
\max_{\zeta<x,y<1-\zeta}( 2uv-u-v-1)<1
\]
where the last inequality follows from
\[
2uv \leq u^2+ v^2 < u+v
\]
for $u,v<1-\zeta <1$.
From inequality (\ref{waf}), we obtain using $d\geq 2$:
\beq
&&\sum_{x\in C_k}\sum_{y\not=x, |y-x|\leq \Delta}\ \prodpee (\theta_y
E_n|\theta_x E_n) \ \prodpee (E_n)
\nonumber\\
&\leq &
(k+1)^d (\Delta+1)^d\sup_{\si,\eta}\sup_{k\geq n}\sup_{x_1,\ldots x_k\in\Z^d} 
\prodpee (\si_{x_1}=\eta_{x_1},\ldots,\si_{x_k}=\eta_{x_k}|(\si,\eta)_{\Z^d\setminus\{x_1,\ldots x_k\}})
\nonumber\\
&\leq & (1-\delta)^n\,.
\nonumber
\eeq
Therefore, choosing $\Delta= n^{d+1}$, we obtain
\be
\nonumber
\sum_{x\in C_k}\sum_{y\not=x, |y-x|\leq \Delta} \prodpee (\theta_y E_n|\theta_x E_n) \prodpee (E_n)
\leq (k+1)^d C
\ee
where
\[
C = \sup_{n} n^{d(d+1)} (1-\delta)^n <\infty\,.
\]
The third term in the decomposition (\ref{wif}) is estimated as in the proof
of Lemma \ref{pospar}. At this point we can repeat the proof of Lemma \ref{pospar}.
\epr

\subsection{Poisson law for occurrences}

For a {\em good pattern} $A=A_n$ supported on $\ce_n$, we
define the second occurrence time by the relation:
\be
\nonumber
(T^2_{A} (\si)\leq k^d) = (N(A,V_k,\si)\geq 2)
\ee
and the restriction that $T^2_{A}$ can only take
values $(k+1)^d$, $k\in\N$.
Similarly we define the $p$-th occurrence time:
\be
\nonumber
(T^p_{A} (\si)\leq k^d) = (N(A,V_k,\si)\geq p)
\ee
and the same restriction. The following proposition
shows that in the limit $n\to\infty$,
properly normalized increments of the process $\{T^k_{A_n}:k\in\N\}$
converge to a sequence of independent exponentials.
This implies convergence of the finite dimensional
distributions of the counting process to a Poisson process
defined in \eqref{croquemonsieur}.
\bp\label{indep}
Let $A_n$ be a good pattern (in the sense of Definition \ref{rouletterusse}).
Define $\tau^p_{A_n}= T^p_{A_n}- T^{p-1}_{A_n}$, where $T^0_{A_n}=0$.
For all $p\in\N$, $t_1,\ldots, t_p\in [0,\infty)$,
$$
\lim_{n\to\infty}
\pee\left(
\left[\tau^p_{A_n} \geq \sum_{i=1}^p t_i/\pee (A_n)\right]
\cap \left[\tau^{p-1}_{A_n} \leq \sum_{i=1}^{p-1} t_i/\pee (A_n)\right]
\cap\ldots\cap\left[ \tau^{1}_{A_n} \leq t_1/\pee (A_n)\right]\right)=
$$
$$
e^{-(t_1+\ldots t_k)}(1-e^{-(t_1+\ldots t_{k-1})})\cdots (1-e^{-t_1})\,.
$$
\ep
\bpr
We start with the case of two occurrence times $T_1, T_2$:
\beq
&&\pee\left( T_1 \leq \frac{t}{\pee (A)} \cap T_2 \geq \frac{s}{\pee (A)}+ T_1 \right)
\nonumber\\
&=&
\sum_{k\leq \frac{t}{\pee (A)}}
\pee \left(T_2 \geq 
\frac{s}{\pee (A)}+ k\ \Big|\ T_1 =k\right)\ \pee (T_1 =k)\,.
\nonumber
\eeq

Let us denote by $\ce_k$ the cube
defined by the relation $(T_1\leq k) = (A\prec \ce_k)$,
and by $A\prec^1 C_k$ the event
that $A$ appears for the first time in $C_k$
(more precisely $A\prec^1 C_k$ abbreviates the event $(T_1 =k)$, i.e.,
$\cap_{l<k}(A\nprec C_l) \cap (A\prec C_k)$).

Let us denote by $\ce_k^\Delta$ the $\Delta$-extension
of $\ce_k$, i.e., the minimal cube containing
$\ce_k$ such that $\partial \ce^\Delta_k$ and $\partial C_k$
are at least $\Delta$ apart. Recall that $C(t/\pee (A))$ denotes
the maximal cube of the form $C_k= [0,k]^d\cap\Z^d$ such that
$|C_k|\leq t/\pee (A)$. Remember that 
\[
\frac{|C(t/\pee (A))|}{t/\pee (A)}\to 1
\]
as $n\to \infty$.

\bl
If $A$ is a good pattern, then we have the estimate
$$
\pee \left( T_2 \geq \frac{s}{\pee (A)} + k\ \Big| \ A\prec^1 \ce_k\right)
-
\pee \Bigl(A\nprec C\left(\frac{s}{\pee(A)}\right)
\setminus \ce^\gap_k\ \Big| \ A\prec^1 \ce_k\Bigr)\leq 
$$
$$
\gap f_A^{(d-1)/d} e^{-c n^d}\,.
$$
\el
\bpr
The proof is identical to that of Lemma \ref{goodlemma}.
\epr
Now we want to replace
\be
\nonumber
\pee \Bigl(A\nprec C\left(\frac{s}{\pee(A)}\right)
\setminus \ce^\gap_k \ \Big|\ A\prec^1 \ce_k\Bigr)
\ee
by
the unconditioned probability of the same event. We make
the choice $\gap = n^{d+1}$.
By the disagreement percolation
estimate, this gives an error which can be
bounded by 
$$
\sum_{k\leq t/\pee(A)}
\pee (T_1=k)\Bigl[\pee \Bigl(A\nprec C\left(\frac{s+t}{\pee(A)}\right)
\setminus \ce^\gap_k \Big| A\prec^1 \ce_k\Bigr)
-
\pee
\Bigl(A\nprec C\left(\frac{s+t}{\pee(A)}\right)
\setminus \ce^\gap_k \Bigr)\Bigr]
\leq
$$
$$
\sum_{k\leq t/\pee(A)}
e^{-c\Delta} \leq t^2 e^{c n^d} e^{-c' n^{d+1}}\,.
$$
Finally, we have
$$
\sup_{k\leq t/\pee (A)}\Bigl[\pee
\Bigl(A\nprec C\left(\frac{s+t}{\pee(A)}\right)
\setminus \ce^\gap_k \Bigr)
-
\pee
\Bigl(A\nprec C\left(\frac{s+t}{\pee(A)}
\right)\setminus
C\left(\frac{t}{\pee(A)}
\right)\Bigr)
\Bigr]\leq 
$$
$$
\Delta (t/\pee (A))^{(d-1)/d} \pee (A)
=\Delta t^{(d-1)/d} \pee(A)^{1/d}\,.
$$
By the exponential law, we have,
using $ |C((t+s)/\pee(A))\setminus C(t/\pee (A))|= t/\pee (A)$:
\be
\nonumber
\pee
\Bigl(A\nprec C\left(\frac{s+t}{\pee(A)}
\right)\setminus
C\left(\frac{t}{\pee(A)}
\right)\Bigr)
= \exp (-\lambda_{A} s) +\epsi_n
\ee
where $\epsi_n= \epsi (n,t,s)\to 0$
as $n\to\infty$. Which gives:
\beq
&&\lim_{n\to\infty}
\left(\pee ( \tau_2 \geq s/\pee (A) \cap \tau_1 \leq t/\pee (A))
-\lim_{n}\pee (\tau_1 \leq t/\pee (A)) e^{-\lambda_A s}\right)
\nonumber\\
&= &
\lim_{n\to\infty}\left(\pee ( \tau_2 \geq s/\pee (A) \cap \tau_1 \leq t/\pee (A))
-(1-e^{-\lambda_A t}) e^{-\lambda_A s}\right) 
=0\,.
\nonumber
\eeq
This proves the statement of the proposition
for $k=2$, the general case is analogous and left
to the reader.
\epr
The following proposition follows immediately from
Proposition \ref{indep}
\bp
Let $A_n\in\gee_n$ be a good pattern supported
on $\ce_n$. Then the finite dimensional marginals
of the process $\{ N^{n}_{t/\lambda_{A_n}} :t\geq 0 \}$
converge to the finite dimensional marginals of a mean
one Poisson process as $n$ tends to infinity.
\ep
In order to obtain convergence in the Skorokhod
space, we have to prove
tightness. This is an immediate consequence of the following simple
lemma for general point processes, applied to
\[
N^n_t = N(A_n, C(t/\pee(A_n)), \si)\,.
\]
\bl
Let $\{ N^n_t :t\geq 0\}$ be a sequence of point processes with
path space measures $\pee^T_n$ on $D([0,T],\N)$. If there exists
$C>0$ such that for all $n$ and for all $t\leq T$ we have
the estimate
\be\label{krak}
\E_n^T (N^n_t) \leq Ct
\ee
then the sequence $\pee^T_n$ is tight.
\el
\bpr 
From (\ref{krak}) we infer for all $n$, $t\leq T$
\[
\pee_n^T ( N^n_t \geq K) \leq C T/K\,.
\]
Hence
\be\label{ping}
\lim_{K\uparrow \infty}\sup_{0\leq t\leq T}\sup_n \pee_n^T (N^n_t \geq K) =0
\ee
For a trajectory $\omega \in D([0,T],\N)$ one defines
the modulus of continuity
\be
\nonumber
w_\gamma (T,\omega)=
\inf_{(t_i)_{i=1}^N} \sup_{i=1}^N |\omega_{t_i}-\omega_{t_{i-1}}|
\ee
where the infimum is taken over all partitions $t_0=0<t_1<\ldots <t_N=t$ such that
$t_i- t_{i-1} \geq\gamma$.
If for some $\epsilon >0$ $w_\gamma (T,\omega)\geq \epsilon$, then the number of jumps of $\omega$ in $[0,T]$ is
at least $[T/\gamma]$.
Hence we obtain
using (\ref{krak}):
\be
\nonumber
\pee_n^T (w_\gamma (T,\omega)\geq \epsilon)\leq \pee_n^T (N^n_T\geq T/\gamma) \leq C\gamma\,.
\ee
This gives for all $\epsi >0$:
\be\label{pong}
\lim_{\gamma\downarrow 0} \sup_n \pee^T_n (w_\gamma(T,\omega) \geq \epsilon ) =0\,.
\ee
Combination of (\ref{ping}) and (\ref{pong}) with the tightness criterion
\cite[p. 152]{Land} yields the result.
\epr

\br
With much more effort, one can obtain precise bounds for the difference
$$
\big| \pee(N^n_t/\lambda_{A_n} = k) - \frac{t^k}{k!}e^{-t}\big|
$$
which are well-behaved in $n$, $t$ and $k$. In particular, from
such bounds one can obtain convergence of all moments of
$N^n_t/\lambda_{A_n}$ to the corresponding Poisson moments.
This is done in \cite{abadipreprint} in the context of
mixing processes.
\er

\subsection{Gumbel law}

For $\eta,\si\in\Omega$ denote
\be
\nonumber
\veek_0 (\eta,\si) = \bigcup \{ G_k : \si_{G_k} = \eta_{G_k}\}\,.
\ee
We start with the following simple lemma:
\bl\label{simpleshit}
\begin{enumerate}
\item
There exists $\delta>0$ such that for all $\eta\in\Omega$:
\be
\nonumber
\inf_{k\in\N} 
\frac{\pee\otimes\pee(\veek_0 \supset G_{k+1})}
{\pee(\veek_0 \supset G_{k})}\geq \delta\,.
\ee
\item
There exists a non-decreasing sequence $u_n\uparrow\infty$
such that for all $n\in\N$:
\be
\nonumber
1\leq n\pee (\veek_0\supset G_{u_n})\leq \frac{1}{\delta}\,\cdot
\ee
\end{enumerate}
\el
\bpr
For item 1:
\beq\label{pincemonseigneur}
\frac{\pee\otimes\pee(\veek_0 \supset G_{k+1})}
{\pee\otimes\pee(\veek_0 \supset G_{k})}
&= &
\pee\otimes\pee (\eta_{x_{n+1}}=\si_{x_{n+1}}|\si_{G_n}=\eta_{G_n})
\nonumber\\
&\geq &
\inf_{\xi,\si} \pee\otimes\pee (\si_x=\eta_x|\si_{\Z^d\setminus\{x\}},\xi_{\Z^d\setminus\{x\}})
\nonumber\\
&=& \delta >0
\nonumber
\eeq
where the last inequality follows from the fact
that $\pee\otimes\pee$ is a Gibbs measures.
For item 2, put
\[
f(n) = \pee\otimes\pee(\veek_0 \supset G_n)
\]
and
\beq
u_n^+ &=& \min\{ k: f(k)\leq 1/n \} 
\nonumber\\
u_n^- &=& \max\{ k: f(k) \geq 1/n \}
\nonumber
\eeq
Clearly, 
\[
u_n^-\leq u_n^+\leq u_n^- + 1\,.
\]
Now choosing $u_n= u_n^-$ and using \eqref{pincemonseigneur}, we obtain
\beq
\frac{1}{n}&\leq & \pee\otimes\pee (\veek_0 \supset G_{u_n})
\nonumber\\
&=&
\frac{\pee\otimes\pee (\veek_0 \supset G_{u_n})}{\pee\otimes\pee
(\veek_0 \supset G_{u_{n}+1})}\ \pee\otimes\pee (\veek_0 \supset G_{u_n+1})\nonumber\\
&\leq & \frac{1}{\delta n}\,\cdot
\nonumber
\eeq
\epr
We now adapt our definition of matching time to the sequence
of sets $G_n$:

\be
\nonumber
\tau_n^\gee (\eta,\si) = \min \{ k: \exists x: G_n+ x\subset G_k
\ \mbox{such that}\ \si_{G_n+x}=\eta_{G_n+x}\}\,.
\ee
We have the relation
\be
\nonumber
(\M_n (\eta,\si)\geq k)= (\tau_k^\gee (\eta,\si)\leq n)\,.
\ee
In words: the maximal matching inside $G_n$ is greater than or equal to $k$ if and only if the
first time that a matching on a set $G_k$ happens is not larger
than $n$.
Now we choose $k=u_n +x$ ($x\in\N$)
and use the exponential law for matching times:
\[
\prodpee( \tau_{u_n}^\gee (\eta,\si)\leq n)= 1- \exp( -\lambda_n \prodpee( \si_{G_{u_n+x}}=\eta_{G_{u_n+x}}))
+\epsilon_n
\]
where $\epsilon_n$ goes to zero as $n$ goes to infinity.
By the choice of $u_n$,
\be\label{dragon}
\prodpee( \si_{G_{u_n+x}}=\eta_{G_{u_n+x}})
=\prodpee (\veek_0 (\eta,\si)\supset G_{u_n +x})\in
\left[\frac{A}{n} e^{-\nu x},\frac{B}{n} e^{-\nu' x}\right] 
\ee
where $A,B\in(0,\infty)$ and
\[
0<e^{-\nu} = \liminf_{n\to\infty} \frac{\prodpee(\si_{G_{n+1}}=\eta_{G_{n+1}})}
{\prodpee(\si_{G_{n}}=\eta_{G_{n}})}<1
\]
and
\[
0<e^{-\nu'} = \limsup_{n\to\infty} \frac{\prodpee(\si_{G_{n+1}}=\eta_{G_{n+1}})}
{\prodpee(\si_{G_{n}}=\eta_{G_{n}})}<1\,.
\]
Here the inequality for the $\liminf$ is an immediate consequence
of Lemma \ref{simpleshit}, and the inequality for the $\limsup$ is derived in a 
completely analogous way, using the Gibbs property. The theorem
now follows immediately from (\ref{dragon}).

\subsection{Proof of Theorem \protect\ref{corollary-OW}}\label{proof-corollary-OW}

We start by showing the following summable upper-bound of
$$
\Pr\{\sigma~:\log(\ret_{\sigma_{\ce_n}}(\sigma) \Pr(\sigma_{\ce_n}))\geq
\log t\}\le
$$
$$
\sum_{A_n\in\gee_n}
\Pr(A_n)\
\Pr\{\sigma~:\log(\ret_{A_n}(\sigma) \Pr(A_n))\geq \log t\ \vert\ A_n\}
+\sum_{A_n\in \gee_n^c}\Pr(A_n)\,.
$$
From Theorem \ref{main2} and Lemma \ref{goodpatterns}
we get for all $0<t< e^{\kappa n^d}$
$$
\Pr\{\sigma:\log(\ret_{\sigma_{\ce_n}}(\sigma) \Pr(\sigma_{\ce_n}))\geq \log
t \}\leq  e^{-c'n^d}+ e^{-\Lambda_1 t} + e^{-cn^d}\, .
$$
Take $t=t_n=\log(n^{\epsilon})$, $\epsilon>\Lambda_1^{-1}$, to get
$$
\Pr\{\sigma:\log(\ret_{\sigma_{\ce_n}}(\sigma) \Pr(\sigma_{\ce_n}))\geq
\log\log(n^{\epsilon})\}
\leq  e^{-c'n^d} + \frac1{n^{\epsilon\Lambda_1}}+  e^{-cn^d}\,.
$$
An application of the Borel-Cantelli lemma leads to
$$
\log\left(\ret_{\sigma_{\ce_n}}(\sigma) \Pr(\sigma_{\ce_n})\right)\leq
\log\log(n^{\epsilon})\quad\textup{eventually a.s.}\,.
$$
For the lower bound first observe that Theorem \ref{main2}
gives, for all $0<t< e^{c n^d}$
$$
\Pr\{\sigma:\log(\ret_{\sigma_{\ce_n}}(\sigma) \Pr(\sigma_{\ce_n}))\leq \log
t\}
\leq  e^{-c'n^d}+1-\exp(-\Lambda_2 t)+ e^{-cn^d}\, .
$$
Choose $t=t_n=n^{-\epsilon}$, $\epsilon>1$, to get, proceeding as before,
$$
\log\left(\ret_{\sigma_{\ce_n}}(\sigma) \Pr(\sigma_{\ce_n})\right)\geq
-\epsilon\log n\quad\textup{eventually a.s.}\,.
$$
Finally, let $\epsilon_0 = \max( \Lambda_1^{-1}, 1)$. 

\subsection{Proof of Theorem \ref{waiting-time}}\label{wtproof}

We first show that the strong approximation formula
\eqref{strong-approximation} holds
with $\weee(\ce_n,\eta,\sigma)$ in place of
$\ret_{\sigma_{\ce_n}}(\sigma)$ with respect to the measure
$\mathbb{Q}\otimes\Pr$.
We have the following identity:
$$
\int d\mathbb{Q}(\eta)\
\Pr\left\{\sigma: \T_{\eta_{\ce_n}}(\sigma)> \frac{t}{\Pr(\eta_{\ce_n})}\right\}=
$$
$$
(\mathbb{Q}\otimes \Pr)\left\{(\eta,\sigma): \weee(\ce_n,\eta,\sigma)>
\frac{t}{\Pr(\eta_{\ce_n})}\right\}\,.
$$
This shows that Theorem \ref{main} remains
valid if we replace $\T_{\eta_{\ce_n}}(\sigma)$ with $\weee(\ce_n,\eta,\sigma)$  
and $\Pr$ with $\mathbb{Q}\otimes\Pr$, hence so is Theorem
\ref{corollary-OW}.
Therefore for $\epsilon$ large enough, we obtain
\be\label{SAW}
-\epsilon\log n
\leq\log(\weee(\ce_n,\eta,\sigma)
  \Pr(\eta_{\ce_n}))
\leq \log\log n^\epsilon
\ee
for $\mathbb{Q}\otimes\Pr$-eventually almost every $(\eta,\sigma)$.
Write
$$
\log(\weee(\ce_n,\eta,\sigma) \Pr(\sigma_{\ce_n}))=
\log\weee(\ce_n,\eta,\sigma)+\log\mathbb{Q}(\eta_{\ce_n})-
\log\frac{\mathbb{Q}(\eta_{\ce_n})}{\Pr(\eta_{\ce_n})}
$$
and use (\ref{SAW}). After division by $n^d$, we obtain (\ref{ASW})
since
$\limn\frac1{n^d}\log\mathbb{Q}(\sigma_{\ce_n})=-s(\mathbb{Q})$,
$\mathbb{Q}$-a.s. by the Shannon-McMillan-Breiman Theorem and
$\limn\frac1{n^d}\log\frac{\mathbb{Q}(\eta_{\ce_n})}{\Pr(\eta_{\ce_n})}=s(\mathbb{Q}|\Pr)$,
$\mathbb{Q}$-a.s. by the Gibbs variational principle (See e.g. \cite{acrv} for a proof). 

\subsection{Proof of Theorem \protect\ref{LD-waiting}}\label{LD-proof}

We follow the line of proof of \cite{acrv} to compute $\mathcal{W}(q)$.
The only extra complication in our case is that the bound
$$
\pee \left( \T_{A_n} > \frac{t}{\pee(A_n)}\right) \leq e^{-ct}
$$
for all $t>0$ cannot be obtained directly from Theorem \ref{main}.
Instead we will use the following lemma which shows that such a bound
can be obtained by a rough version of the iteration lemma.
Given this result, the proof of \cite{acrv} can be repeated.

\bl
\ben
\item There exists $c>0$ such that for all patterns $A_n\in \{+,-\}^{\ce_n}$
$$
\pee \left( \T_{A_n} > \frac{t}{\pee(A_n)}\right) \leq e^{-ct}\,.
$$
\item There exists $\delta\in(0,\frac1{2})$ such that for all $n$
and all pattern $A=A_n$
$$
0<\delta<\pee ( T_A > \frac{1}{2\pee(A)})<1-\delta<1\,.
$$
\een
\el

\bpr
To prove the first inequality, we fill part of the cube
$C(t/\pee(A))$ with little cubes of size $f_A$ (where $f_A$ is
defined in Lemma \ref{marx}), with $k\geq t/(2\pee(A)f_A)$.
The gaps $\Delta$ separating the different cubes are taken 
equal to $\lceil t n^{d+1}\rceil$.
We then have the following
\[
\pee( T_A > t/\pee (A))\leq \pee (A\nprec \cup_{i=1}^K C_i)\,.
\]
Notice that we do not have to estimate here the
probability that the pattern is not in the gaps since
we only need an upper bound.
Now 
\[
\alpha_K=\pee ( A\nprec  \cup_{i=1}^K C_i)= 
\]
\[
\pee (A\nprec C_1|A\nprec\cup_{i=2}^K C_i)
\pee(A\nprec\cup_{i=2}^K C_i)= \pee (A\nprec C_1|A\nprec\cup_{i=2}^K C_i)\alpha_{K-1}\,.
\]
Using the disagreement percolation estimate, we have
$$
\pee (A\nprec C_1|A\nprec\cup_{i=2}^K C_i)- \alpha_1 \leq e^{-\Delta}\,.
$$
Therefore
\[
\alpha_{K}\leq \alpha_{K-1}\alpha_1 + e^{-\Delta}\,.
\]
Iterating this inequality gives, using $\Delta=\lceil t n^{d+1}\rceil$,
\[
\alpha_K \leq \alpha_1^K + e^{-tn^{d+1}}e^{c n^d} t
\]
Now we use $K> t/2\pee(A) f_A$, and Lemma \ref{pospar}
to obtain:
\[
\alpha_K \leq (1-\la_1 f_A \pee (A))^{t/(2f_A\pee(A))} + e^{-ct}
\]
which implies the first inequality of the lemma.

The second inequality follows directly from Lemma \ref{pospar}.
\epr

\subsection{Proof of Theorem \protect\ref{matching-time}}\label{pape}

The proof of \eqref{vampire} is identical to the proof of \eqref{dracula} but
using the exponential law for the matching time.
Formula \eqref{ASM} follows from 
$$
\pee\!\otimes\!\pee(\si_{\ce_n}=\eta_{\ce_n})= \sum_{\si_{\ce_n}\in\{+,-\}^{\ce_n}} \pee(\si_{\ce_n})^2
$$
and the definition of $\mathcal{W}$.


\end{document}